\documentclass[preprint,aps]{revtex4}
\newcommand {\bea}{\begin{eqnarray}}
\newcommand {\eea}{\end{eqnarray}}
\newcommand {\be}{\begin{equation}}
\newcommand {\ee}{\end{equation}}

\usepackage{graphicx}
\usepackage{amsmath}
\usepackage{bm}
\usepackage{pslatex}
\usepackage{eucal}
\usepackage[dvips]{color}
\def\){\right)} 
\def\({\left(} 
\def\]{\right]} 
\def\[{\left[}

\begin{document}


\title{Density Functional Theory for non-relativistic Fermions in the \\
Unitarity Limit}

\author{Gautam Rupak\footnote{present address: Department of Physics 
and Astronomy, Mississippi State University, Mississippi State, MS 39762.
email: grupak@gmail.com} 
and Thomas Sch\"afer\footnote{email: thomas\_schaefer@ncsu.edu}}

\affiliation{Department of Physics, North Carolina State University,
Raleigh, NC 27695}

\begin{abstract}
We derive an energy density functional for non-relativistic spin 
one-half fermions in the limit of a divergent two-body scattering 
length. Using an epsilon expansion around $d=4-\epsilon$ spatial 
dimensions we compute the coefficient of the leading correction 
beyond the local density approximation (LDA). In the case of $N$ 
fermionic atoms trapped in a harmonic potential this correction 
has the form $E=E_{LDA}(1+c_s (3N)^{-2/3})$, where $E_{LDA}$ is 
the total energy in LDA approximation. At next-to-leading order 
in the epsilon expansion we find $c_s=1.68$, which is significantly 
larger than the result for non-interacting fermions, $c_s=0.5$.
\end{abstract}

\maketitle

\newpage

\section{Introduction}
\label{sec_intro}

 Density functional theory (DFT)~\cite{Hohenberg:1964,Dreizler:1990}
is widely used in condensed matter physics and quantum chemistry to 
treat quantum many-body systems. DFT relies on the fact that one can 
demonstrate the existence of an energy functional that depends only on 
the fermion density, and not on the wave functions, that has the
property that the ground state energy can be obtained by minimizing
the functional with respect to the density. There is renewed interest
among nuclear theorists in developing a ``Universal Nuclear Energy Density 
Functional'' \cite{unedf} which describes nuclear properties all across 
the isotopic chart, including very neutron rich nuclei far from the 
valley of stability. Ideally, this functional should be derived from 
a systematic theory of the nucleon-nucleon interaction. A modern approach 
to nuclear forces is provided by the effective field theory (EFT) method 
\cite{Weinberg:1990rz,Beane:2000fx,Bedaque:2002mn,Furnstahl:2008df}.
EFT starts from the most general local lagrangian containing nucleon 
and pion fields that respects the basic symmetries of QCD. A power
counting scheme determines the number of terms in the lagrangian,
as well as the number of diagrams, that have to be kept at any 
given order in a low energy expansion. 

 If the effective field theory is perturbative then there are 
systematic methods for determining the energy density functional
\cite{Negele_1972,Puglia:2002vk}. However these methods are very
cumbersome if long-range forces, such as pion exchange or collective
modes, are important. In the case of non-perturbative EFT's no 
systematic calculations of the energy density functional exist. 
In nuclear matter, and in cold fermionic gases near a Feshbach 
resonance, an important source of non-perturbative physics
is the large two-body scattering length. 

 In the present work we try to address some of these problems
by studying the energy density functional for a dilute system of
non-relativistic spin 1/2 fermions with an infinite two-body 
scattering length. Because the s-wave cross section saturates
the unitarity bound this limit is often referred to as the ``unitarity
limit''. The energy density functional in this limit is important 
for the study of  neutron star crusts and neutron halos in nuclei. 
It can also be used to describe trapped fermionic atoms in the 
vicinity of a Feshbach resonance. The Fermi gas in the unitarity
limit exhibits a number of interesting non-perturbative phenomena. 
It is a superfluid, and the ratio of the gap over the Fermi energy 
is large. Superfluidity implies that the $U(1)$ phase symmetry is 
spontaneously broken and the low energy or momentum response is 
carried by Goldstone modes. 

  We shall compute the energy density functional up to next-to-leading
order (NLO) in an expansion in derivatives of the density. Our procedure 
is based on an effective lagrangian for the Fermi gas in the unitarity
limit derived in \cite{Son:2005rv}. We will determine the coefficients
in this lagrangian using an epsilon expansion around $d=4-\epsilon$ 
spatial dimensions \cite{Nussinov:2004,Nishida:2006br}. As a by-product 
we compute the phonon dispersion relation and the static susceptibility 
at NLO in the epsilon expansion. Our result for the energy density 
functional is rigorous when applied to infinite systems in which the 
density varies smoothly, but there are some limitations in the case 
of finite systems with a sharp surface, such as fermions confined in 
a harmonic trap. Pairing leads to an odd-even effect in the dependence
of the energy on the number of particles, which a local energy density 
functional that depends only on the particle density cannot describe. 
Also, the gradient expansion breaks down near the surface of the 
system and the expansion of the energy in inverse fractional powers 
of the number of particles cannot be pushed to arbitrarily high 
order \cite{Son:2005rv}. We shall discuss some possible approaches 
to overcome these limitations in Sec.~\ref{sec_out}.

\section{Effective lagrangian and energy density functional}
\label{sec_eft}

 The energy density functional describes the response of the 
system to smooth variations in the density. This functional can
be related to the effective lagrangian that governs the response 
to slowly varying external fields. The effective Lagrangian 
at NLO in derivatives of the external potential is \cite{Son:2005rv}
\be
\label{l_eft}
  {\cal L} = c_0 m^{3/2} X^{5/2} 
  + c_1 m^{1/2} \frac{(\vec{\nabla} X)^2}{\sqrt{X}} 
  + \frac{c_2}{\sqrt m} 
     \left[ \left(\nabla^2\varphi\right)^2 
           - 9m \nabla^2 V\right] \sqrt X\,,
\ee
where we have defined 
\be
  X = \mu - V - \dot\varphi 
    - \frac{(\vec{\nabla}\varphi)^2}{2m}\,.
\ee
The lagrangian contains the Goldstone boson (phonon) field $\varphi
(\vec{x},t)$, the chemical potential $\mu$, and the external potential 
$V(\vec{x},t)$. The mass of the fermion is denoted by $m$. The functional 
form of the effective lagrangian is fixed by the symmetries of the problem, 
Galilean invariance, $U(1)$ symmetry, and conformal symmetry. The NLO 
effective lagrangian is characterized by three dimensionless parameters, 
$c_0,c_1,c_2$. These parameters can be related to physical properties 
of the system. The first parameter, $c_0$, can be related to the equation 
of state. We have 
\be
  c_0 = \frac{2^{5/2}}{15\pi^2\xi^{3/2}}\,,
\ee
where $\xi$ determines the chemical potential in units of the 
Fermi energy, $\mu=\xi\epsilon_F$ with $\epsilon_F=k_F^2/(2m)$. 
The two NLO parameters $c_1,c_2$ are related to the momentum 
dependence of correlation functions. The phonon dispersion
relation, for example, is given by \cite{Son:2005rv}
\be
  q_0 =  v_s q\left[ 1
   - \pi^2\sqrt{2\xi}\left(c_1+\frac{3}{2} c_2\right)
                \frac{q^2}{k_F^2} + O(q^4\log(q^2))\right]
\ee
where $v_s=\sqrt{\xi/3} v_F$ is the speed of sound and $v_F=k_F/m$
is the Fermi velocity. The static susceptibility is defined by
\be
 \chi(q) = -i\int dt\,d^3x\; e^{-i\vec{q}\cdot\vec{x}}\,
   \langle \psi^\dagger\psi(0) \psi^\dagger\psi(t,\vec{x})\rangle ,
\ee
where $\psi^\dagger\psi\equiv\psi^\dagger_\alpha\psi^\alpha$
($\alpha=1,2$) is the sum of the spin up and down densities.
The susceptibility is related to a different linear combination 
of $c_1$ and $c_2$ \cite{Son:2005rv}, 
\be
  \chi(q) = - \frac{mk_F}{\pi^2\xi} \left[
    1 + 2\pi^2\sqrt{2\xi}\left(c_1 - \frac{9}{2} c_2\right) 
    \frac{q^2}{k_F^2}+O(q^4\log(q^2))\right] .
\ee
The effective lagrangian can be used to compute the groundstate
energy of fermions confined by an external potential. The 
energy of $N$ fermions in a spherically symmetric trap 
$V(x)=\frac{1}{2}m\omega^2x^2$ is 
\be 
E=\frac{\sqrt{\xi}}{4} \omega (3N)^{4/3}
 - 3\sqrt{2}\pi^2\xi\omega \left(c_1-\frac{9}{2}c_2\right)
   (3N)^{2/3} + \ldots \, .
\ee
In this work we will derive an energy functional that depends on 
the local density $n(x)$. This functional is the Legendre transform 
of the pressure, 
\be 
{\cal E}[n(x)] = \mu n(x)-P[\mu-V(x)] . 
\ee
The energy functional is easily derived from the effective lagrangian.
Up to NLO in the derivative expansion it is sufficient to consider 
the tree-level effective lagrangian \cite{Son:2005rv}. The only difficulty 
is to invert the relationship between the density and and the chemical 
potential. This can be done order by order in the derivative expansion. 
We write
\begin{eqnarray}
n[\mu-V(x)] &=& n_0[\mu-V(x)]+\delta\, n_1[\mu-V(x)]
  +\delta^2\, n_2[\mu-V(x)]+ \ldots \\
\mu-V(x)    &=& \mu_0[n(x)]  +\delta\, \mu_1[n(x)] 
  +\delta^2\, \mu_2[n(x)] + \ldots \\
 {\cal E}[n(x)]&=& {\cal E}_0[n(x)] + \delta\, {\cal E}_1[n(x)]
 + \delta^2\, {\cal E}_2[n(x)]+\ldots ,
\end{eqnarray}
where $\delta$ is used as an expansion parameter. The functions $n_0,n_1,
\ldots$ arise from differentiating the leading order, next-to-leading, 
etc.~terms in the effective lagrangian with respect to $\mu$. The 
functions $\mu_0,\mu_1,\ldots$ can be found by inverting this 
relationship order by order. We find
\be 
{\cal E}_0[n(x)]=n(x)V(x)+\mu_0[n(x)]n(x)-P_0[\mu_0[n(x)]], 
\ee
with $\mu_0[n(x)] = (n_0)^{-1}[n(x)]$ and ${\cal E}_1[n(x)]=
-P_1(\mu_0[n(x)])$. This yields
\be
\label{edf}
{\cal E}(x) = n(x)V(x) 
  + \frac{3\cdot 2^{2/3}}{5^{5/3}mc_0^{2/3}}n(x)^{5/3}
 - \frac{4}{45}\frac{2c_1-9c_2}{mc_0} 
     \frac{\left(\nabla n(x)\right)^2}{n(x)}
 - \frac{12}{5}\frac{c_2}{mc_0} \nabla^2 n(x) \, .
\ee
The first two terms correspond to the local density approximation
(LDA) and the terms proportional to $c_1$ and $c_2$ are the leading 
correction to the LDA involving derivatives of the density. We note
that the last term proportional to $\nabla^2 n(x)$ does not contribute 
to the total energy of a finite system.

\section{Epsilon Expansion}
\label{sec_eps}
\subsection{Lagrangian and Feynman rules}
\label{sec_lag}

 At unitarity the determination of $c_1$ and $c_2$ is a non-perturbative
problem, and we will perform the calculation using an expansion around
$d=4-\epsilon$ spatial dimensions \cite{Nussinov:2004,Nishida:2006br}.
The epsilon expansion has proven to be useful in calculating the 
equation of state \cite{Arnold:2006fr}, the critical temperature
\cite{Nishida:2006rp}, few-body scattering observables \cite{Rupak:2006jj}, 
and the phase structure of spin-polarized systems \cite{Rupak:2006et}.
Our starting point is the lagrangian 
\be
{\cal L}= \Psi^\dagger\left[
     i\partial_0+\sigma_3\frac{\vec\nabla^2}{2m}\right]\Psi
  + \mu\Psi^\dagger\sigma_3\Psi
  +\left(\Psi^\dagger\sigma_+\Psi\phi + h.c. \right)
  -\frac{1}{C_0}\phi^\dagger\phi\ ,
\ee
where $\Psi=(\psi_\uparrow,\psi_\downarrow^\dagger)^T$ is a two-component 
Nambu-Gorkov field, $\sigma_i$ are Pauli matrices acting in the Nambu-Gorkov 
space, $\sigma_\pm=(\sigma_1\pm i\sigma_2)/2$, $\phi$ is a complex boson
field, and $C_0$ is a coupling constant. In dimensional regularization 
the fermion-fermion scattering length becomes infinite for $1/C_0\to 0$. 

 The epsilon expansion is based on the observation that the fermion-fermion 
scattering amplitude near $d=4$ dimensions is saturated by the propagator 
of a boson with mass $2m$. The coupling of the boson to pairs of fermions
is given by
\be
   g  =\frac{\sqrt{8\pi^2\epsilon}}{m}
       \left(\frac{m\phi_0}{2\pi}\right)^{\epsilon/4} \, .
\ee
In the superfluid phase $\phi$ acquires an expectation value $\phi_0=
\langle\phi\rangle$. We write the boson field as $\phi = \phi_0 + g\varphi$. 
The lagrangian is split into a free part 
\be
{\cal L}_0 = \Psi^\dagger\left[i\partial_0+\sigma_3\frac{\vec\nabla^2}{2m}
     + \phi_0(\sigma_{+} +\sigma_{-})\right]\Psi
     + \varphi^\dagger\left(i\partial_0
        + \frac{\vec\nabla^2}{4m}\right)\varphi\, ,
\ee
and an interacting part ${\cal L}_I+{\cal L}_{ct}$, where  
\bea
{\cal L}_I &=& g\left(\Psi^\dagger\sigma_+\Psi\varphi + h.c\right)
     + \mu\Psi^\dagger\sigma_3\Psi  +2\mu \varphi^\dagger\varphi \, , \\
{\cal L}_{ct} &=& 
     - \varphi^\dagger\left(i\partial_0
        + \frac{\vec\nabla^2}{4m}\right)\varphi
     -2\mu \varphi^\dagger\varphi\, . 
\eea
Note that the leading self energy corrections to the boson propagator
generated by the interaction term ${\cal L}_I$ cancel against the 
counterterms in ${\cal L}_{ct}$. The chemical potential term for 
the fermions is included in ${\cal L}_I$ rather than in ${\cal L}_0$. 
This is motivated by the fact that near $d=4$ the system reduces to 
a non-interacting Bose gas and $\mu\to 0$. We will count $\mu$ as a 
quantity of $O(\epsilon)$. The Feynman rules are quite simple. The 
fermion and boson propagators are
\bea
\label{eps_prop}
G(p_0,p) &=& \frac{i}{p_0^2-E_{p}^2}
\left[\begin{array}{cc}
    p_0+\epsilon_{p} &  -\phi_0\\
    -\phi_0        & p_0-\epsilon_{p}
\end{array}\right]  \ ,\\
D(p_0, p) &=& \frac{i}{p_0-\epsilon_{p}/2}\ , 
\eea 
where $E_p^2=\epsilon_p^2+\phi_0^2$ and $\epsilon_p=p^2/(2m)$. The 
fermion-boson vertices are $ig\sigma^\pm$. Insertions of the 
chemical potential are $i\mu\sigma_3$. Both $g^2$ and $\mu$ are 
corrections of order $\epsilon$. 

We shall make use of the following results that have been obtained
at NLO in the epsilon expansion \cite{Nishida:2006br}
\bea 
\label{phi0_eps}
\phi_0 &=& \frac{2\mu}{\epsilon}\,\left[ 1 + 
    (3C-1+\log(2))\,\epsilon + O(\epsilon^2) \right] , \\
\label{n_eps}
 n  &=& \;\frac{1}{\epsilon}\; \left[ 1 - \frac{1}{4}
  \left( 2\gamma-1-2\log(2) \right) + O(\epsilon^2) \right]  
 \left(\frac{m\phi_0}{2\pi}\right)^{d/2}, \\
\label{xi_eps}
\xi &=& \frac{\epsilon^{3/2}}{2}  \left[ 1  
      + \frac{1}{8}\epsilon \log(\epsilon)
      - \frac{1}{4}\left(12C-5+5\log(2)\right)\epsilon 
      + O(\epsilon^2) \right] .
\eea
Here, $\phi_0$ is the expectation value of the boson field, $n$
is the density, and $\xi$ determines the chemical potential in 
units of the Fermi energy, $\mu=\xi \epsilon_F$. The quantity
$C\simeq 0.14424$ is a numerical constant that appears in the 
calculation of the two-loop effective potential, and $\gamma
\simeq 0.57722$ is the Euler constant.

\begin{figure}
\includegraphics[width=14cm]{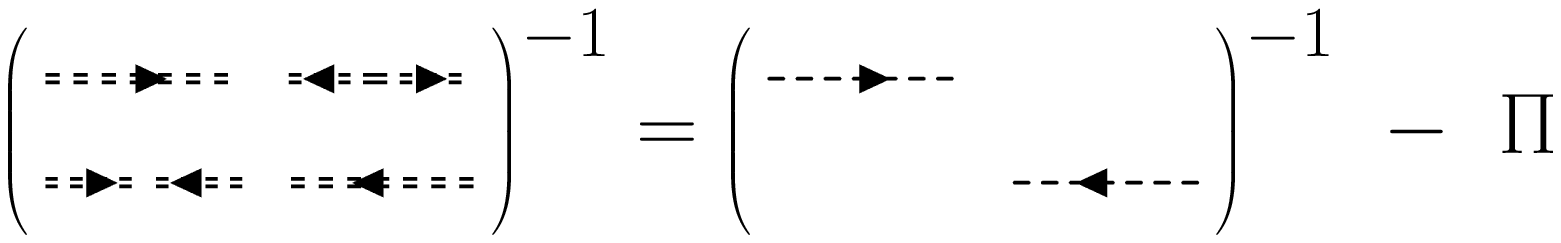}
\caption{Dyson Schwinger equation for the phonon propagator. Dashed 
lines denote the free boson propagator, double dashed lines denote 
the full propagator and $\Pi$ is the boson self energy. Arrows
show the order in which $\varphi$ and $\varphi^*$ are contracted. }
\label{fig_ds}
\end{figure}

\subsection{Phonon propagator}
\label{sec_pho}

\begin{figure}
\includegraphics[width=12cm]{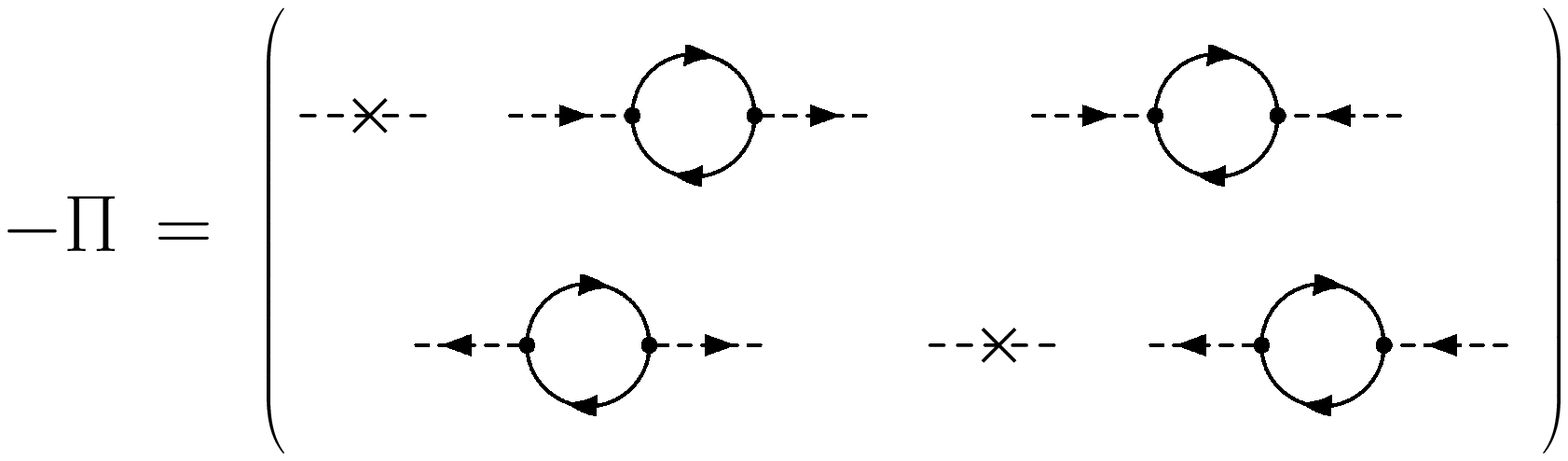}
\caption{Leading order contributions to the boson self energy. Full 
lines denote fermion propagators in the Nambu-Gorkov representation.
Arrows indicate the order in which $\Psi$ and $\Psi^\dagger$ are
contracted. A cross denotes a $\mu$-insertion from ${\cal L}_I$. 
There is a contribution from the first counterterm in ${\cal L}_{ct}$
which is not shown here.}
\label{fig_self}
\end{figure}

 The phonon dispersion relation at LO in the epsilon expansion was
obtained by Nishida in \cite{Nishida:2006rp}. Here, we briefly 
review his results. We introduce a two-component scalar field 
$\Phi=(\varphi,\varphi^*)$. The scalar propagator is
\be 
{\cal D}^{-1}(p) = {\cal D}_0^{-1}(p)-\Pi(p) = 
\left(\begin{array}{cc}
 [{\cal D}^{-1}(p)]_{11} &  [{\cal D}^{-1}(p)]_{12}  \\[0.1cm]
 [{\cal D}^{-1}(p)]_{21} &  [{\cal D}^{-1}(p)]_{22}
 \end{array}\right) ,
\ee
where ${\cal D}$ is the full propagator, ${\cal D}_0$ is the
free propagator, and $\Pi$ is the self energy, see Fig.~\ref{fig_ds}.
The free propagator does not have off-diagonal (anomalous) 
components. The diagonal terms are
\be
 [{\cal D}_0^{-1}(p)]_{11} = [{\cal D}_0^{-1}(-p)]_{22}
 = p_0 -\frac{\epsilon_p}{2} \, .
\ee
The self energy diagram at LO in the epsilon expansion are shown 
in Fig.~\ref{fig_self}. We find
\bea 
\Pi_{11} &=& \Pi_{22} = -2\mu +\frac{3\epsilon\phi_0}{2} 
  + O(\epsilon^2) \, , \\
\Pi_{12} &=& \Pi_{21} = \frac{\epsilon\phi_0}{2} 
  + O(\epsilon^2) \, . 
\eea
At leading order $\mu=\epsilon\phi_0/2$ and 
\be 
{\cal D}(p)= \frac{1}{p_0^2-\frac{\epsilon_p}{2}
                     (\frac{\epsilon_p}{2}+2\mu)}
\left(\begin{array}{cc}
p_0+\frac{\epsilon_p}{2}+\mu & -\mu \\
-\mu  & -p_0+\frac{\epsilon_p}{2}+\mu
\end{array}\right)\, . 
\ee
The dispersion relation is 
\be 
 p_0 = \frac{1}{2}\sqrt{\epsilon_p(\epsilon_p+4\mu)}
     \simeq \sqrt{\mu\epsilon_p} 
       \left( 1 +\frac{\epsilon_p}{8\mu} + \ldots \right) , 
\ee
which shows that the spectrum contains a Goldstone mode with 
a linear dispersion relation, $p_0\simeq v_s p$, where $v_s=
\sqrt{\mu/(2m)}$. 

\subsection{Static susceptibility}
\label{sec_sus}

\begin{figure}
\begin{center}
\raisebox{2cm}{a)}
\includegraphics[width=4cm]{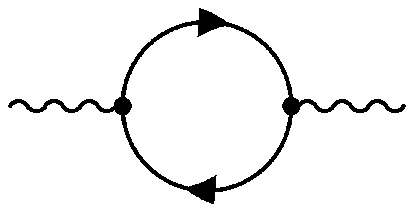}\hspace*{0.5cm}
\raisebox{2cm}{b)}
\includegraphics[width=8cm]{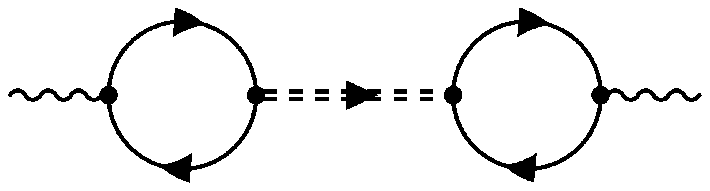}
\end{center}

\vspace*{0.6cm}
\begin{center}
\raisebox{2cm}{c)}\includegraphics[width=9cm]{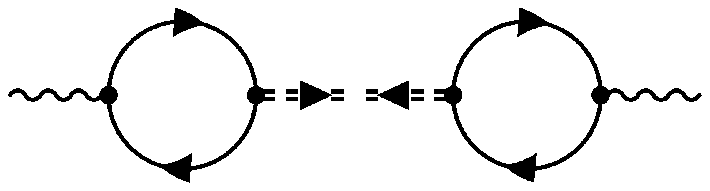}
\end{center}
\caption{Fig.~a) shows the one-loop contribution to the static 
susceptibility. The wavy line denotes an external source coupled 
to $\psi^\dagger\psi$. Figs.~b) and c) show the leading order phonon 
contribution. The double dashed line is full phonon propagator 
defined in Fig.~\ref{fig_ds}.  }
\label{fig_sus}
\end{figure}

 The one-loop contribution to the static susceptibility (see 
Fig.~\ref{fig_sus}a) is 
\bea 
 \chi(q) &=&-i\int \frac{d^dk}{(2\pi)^d}\int \frac{dk_0}{2\pi}
  \; {\rm Tr}\left[ G(k+q/2)\sigma_3 G(k-q/2)\sigma_3\right]  \\
  &=& -\frac{1}{\phi_0} \left\{ \; 1 \;
 - \frac{1}{2} \left( \gamma-1+\log(2) \right) \epsilon
 - \frac{1}{12} \left(\frac{q^2}{m\phi_0}\right)
    + O(\epsilon^2)
 \right\} \left(\frac{m\phi_0}{2\pi}\right)^{d/2}, \nonumber
\eea
where we have expanded $\chi(q)$ in powers of momentum, treating 
$q^2$ as a quantity of order $\epsilon$. We observe that the one-loop
contribution scales as $\chi(0)\sim \epsilon^0$. This should be 
compared to the thermodynamic result $\chi(0)=-(\partial n)/(\partial 
\mu)\sim \epsilon^{-2}$. In order to get an enhancement by two
inverse powers of $\epsilon$ we need to consider graphs that 
contain massless particles. The dominant contribution comes 
from phonons, see Fig.~\ref{fig_sus}b,c. The LO phonon term is 
\be 
\chi(q) =g^2\Big\{ \Pi_{3+}(q){\cal D}_{11}(q)\Pi_{3-}(q)
  + \Pi_{3+}(q){\cal D}_{12}(q)\Pi_{3+}(q) + {\it h.c}\Big\}
\ee
where ${\cal D}_{ij}$ is the phonon propagator and 
\bea
 \Pi_{3\pm}(q) &=&-i\int \frac{d^dk}{(2\pi)^d}\int \frac{dk_0}{2\pi}
  \; {\rm Tr}\left[ G(k+q/2)\sigma_3 G(k-q/2)\sigma_\pm\right] \\
 &=& -\frac{1}{\epsilon\phi_0} \left\{ \; 1 \;
 - \frac{1}{2} \left( \gamma-\log(2) \right) \epsilon
 + \frac{1}{8} \left( \gamma-\log(2) \right)^2 \epsilon^2
 - \frac{1}{24} \left(\frac{q^2}{m\phi_0}\right)\epsilon
 \right\} \left(\frac{m\phi_0}{2\pi}\right)^{d/2} \nonumber
\eea
Using the leading order phonon propagator derived in the 
previous section we find in the static limit
\be 
{\cal D}_{11}(q)+{\cal D}_{12}(q) 
=-\frac{1}{2\mu+\epsilon_q/2} 
=-\frac{1}{2\mu}\left\{ 1-\frac{1}{8}  
   \left(\frac{q^2}{m\mu}\right) + O(q^4) \right\}
\ee

We can now determine the static susceptibility 
\be
\label{chi_phonon_lo}
 \chi(q) = -\frac{2}{\epsilon\mu} \bigg[ 1 
 -  \left( \gamma-\log(2) \right) \epsilon
 + O(\epsilon^2) \bigg]
  \left\{ 1 - \frac{1}{8}\left(\frac{q^2}{m\mu}\right) + O(q^4)
 \right\} \left(\frac{m\phi_0}{2\pi}\right)^{d/2} \, .
\ee
We can compare the momentum independent term to the prediction 
from the relation $\chi(0)=-(\partial n)/(\partial \mu)$. Using 
equ.~(\ref{n_eps}) and (\ref{phi0_eps}) we find, at NLO in the 
$\epsilon$ expansion, 
\be 
\label{chi_sr}
\chi(0)= -\frac{2}{\epsilon\mu} \left\{ \; 1 \;
 - \;\frac{1}{2} \left( \gamma-\log(2) \right) \epsilon
 + O(\epsilon^2) \right\} 
\left(\frac{m\phi_0}{2\pi}\right)^{d/2} \, .
\ee
which agrees at leading order, but not at NLO.

\subsection{Higher order corrections}
\label{sec_nlo}

\begin{figure}
\raisebox{2cm}{a)}
\includegraphics[width=4cm]{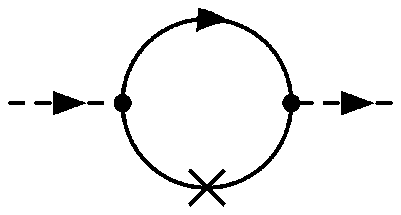}
\hspace*{0.3cm}
\raisebox{2cm}{b)}
\includegraphics[width=4cm]{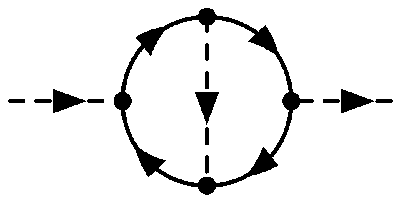}
\hspace*{0.3cm}
\raisebox{2cm}{c)}
\includegraphics[width=4cm]{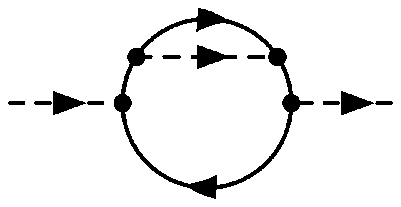}
\caption{Higher order contributions to the diagonal boson self energy
$\Pi_{11}$. Fig.~a) shows a $\mu$-insertion into the one-loop 
self energy. This diagram is combined with the second counterterm from 
${\cal L}_{ct}$ (not shown). Figs.~b) and c) show the ``vertex-type'' 
and ``self energy-type'' two-loop contributions. The corresponding 
contributions to the off-diagonal self energy $\Pi_{12}$ are not 
shown.  }
\label{fig_self_nlo}
\end{figure}

 The LO phonon dispersion relation and susceptibility depend on 
$O(\epsilon)$ terms in the boson self energy. The NLO phonon 
dispersion relation requires $O(\epsilon^2)$ corrections. Since
the LO curvature term in the dispersion relation is proportional
to $p^2/(m\mu)$ we can count $p_0,\epsilon_p$ as quantities of 
order $\epsilon$. 
The one-loop self energies, expanded to NLO in $\epsilon$, are 
given by 
\bea
\Pi_{11} &=& -\left(p_0-\frac{\epsilon_p}{2}\right) 
  \left\{ 1 - \frac{1}{2}\left(\gamma-\log(2)\right)\epsilon \right\}
  +\frac{3\epsilon\phi_0}{2} 
  \left\{ 1 +\frac{1}{6} \left(5-3\gamma-\log(8)\right)\epsilon
   \right\} 
  + \ldots  \\
\Pi_{12} &=& \frac{\epsilon\phi_0}{2}
  \left\{ 1 +\frac{1}{2} \left(1-\gamma-\log(2)\right) \epsilon
   \right\} 
   - \frac{\epsilon\epsilon_p}{8} + \ldots
\eea
and the NLO expression for $\phi_0$ is given in equ.~(\ref{phi0_eps}).
The term $-(p_0-\epsilon_p/2)$ is canceled by the first counterterm
in ${\cal L}_{ct}$. The $\mu$ insertion into the one-loop self energy 
is 
\bea
\Pi_{11} &=& -2\mu \left\{ 1 -\frac{1}{4}\left(
  1+2\gamma-2\log(2)\right)\epsilon + O(\epsilon^2) \right\}
  \\
\Pi_{12} &=& \,\;\frac{\mu\epsilon}{2}\;
    \bigg\{ 1 +  O(\epsilon) \bigg\}  \, , 
\eea
where the term $-2\mu$ is canceled by the second counterterm in 
${\cal L}_{ct}$. There are two two-loop self energy diagrams, see 
Fig.~\ref{fig_self_nlo}. We will compute these diagrams in 
App.~\ref{sec_app}. The result can be written as
\be 
\Pi_{11} = C_1\phi_0 \epsilon^2 , \hspace{1cm}
\Pi_{12} = C_2\phi_0 \epsilon^2 ,
\ee
where $C_{1,2}$ are numerical constants. These two constants are 
constrained by some general relations. First, the phonon is a 
Goldstone mode and the dispersion relation has to satisfy $\omega_q
(q=0)=0$. We also know that the velocity of sound is related to
the equation of state. In $d=4-\epsilon$ dimensions
\be 
 v_s = \sqrt{\frac{\xi}{d}} v_F 
     = \sqrt{\frac{\mu}{2m}}\left( 1+\frac{\epsilon}{8}+O(\epsilon^2)
        \right).
\ee
These two conditions determine $C_{1,2}$. We find
\be
\label{C_12}
 C_1 =-\frac{9}{2}\, C \simeq -0.64908, \hspace{0.5cm}
 C_2 =-\frac{3}{2}\, C \simeq -0.21636.
\ee
In App.~\ref{sec_app} we demonstrate that these results agree
with an explicit calculation of the two-loop self energies. 
With these results, the inverse boson propagator at NLO takes on a 
very simple form. We find
\be 
{\cal D}^{-1} =  Z \left( 
\begin{array}{cc}
p_0-\frac{\epsilon_p}{2} -\mu & 
\frac{\epsilon\epsilon_p}{8} -\mu \\
\frac{\epsilon\epsilon_p}{8} -\mu & 
-p_0-\frac{\epsilon_p}{2} -\mu
\end{array}\right),
\hspace{0.5cm}
Z=1 - \frac{1}{2}\left(\gamma-\log(2)\right)\epsilon 
\ee
The phonon dispersion relation is 
\be 
\label{ph_disp_nlo}
 p_0 = \sqrt{\mu\epsilon_p} 
       \left(1+\frac{\epsilon}{8}\right) 
       \left\{ 1 +\frac{\epsilon_p}{8\mu} 
      \left(1-\frac{\epsilon}{4} \right)+ \ldots \right\}
\ee
 
\begin{figure}
\begin{center}
\raisebox{2cm}{a)}
\includegraphics[width=4cm]{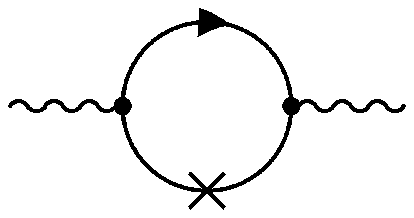}\hspace*{0.5cm}
\raisebox{2cm}{b)}
\includegraphics[width=8cm]{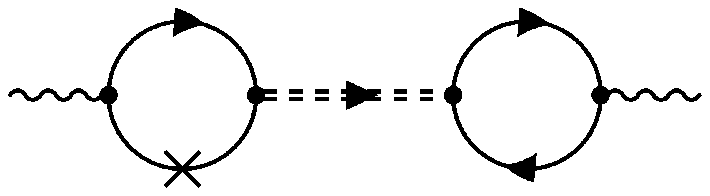}
\end{center}

\vspace*{0.6cm}
\begin{center}
\raisebox{2cm}{c)}
\includegraphics[width=9cm]{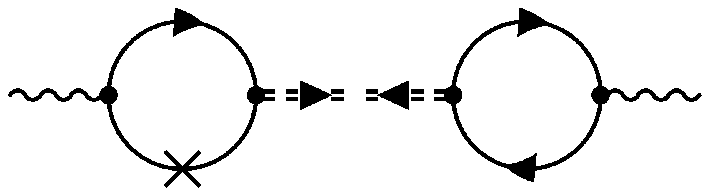}
\end{center}
\caption{Higher order contributions to the static susceptibility.
Fig.~a) shows the $\mu$-insertion into the one-loop diagram, 
Figs.~b) and c) show $\mu$-insertions into the phonon contribution. }
\label{fig_sus_nlo}
\end{figure}

 We now consider NLO corrections to the static susceptibility, see
Fig.~\ref{fig_sus_nlo}. The first diagram is the $\mu$-insertion
into the one-loop graph. This graph is $O(\epsilon\mu)$, which 
is an $O(\epsilon^4)$ correction to the leading order term in 
$\chi(0)$. The second and third diagram contain $\mu$-insertions
into $\Pi_{3+}$. These diagrams are $O(1)$, an $O(\epsilon^2)$ 
correction to the leading term. Two-loop corrections to $\Pi_{3+}$
are also suppressed by at least two powers of $\epsilon$. This
means that, in addition to NLO corrections from the one-loop
$\Pi_{3+}$ already given in equ.~(\ref{chi_phonon_lo}), NLO corrections 
to the static susceptibility arise solely from higher order corrections 
to the boson propagator. We find 
\be 
\label{chi_nlo}
\chi(q) = -\frac{2}{\epsilon\mu} 
    \left[ 1 - \frac{1}{2}\left(\gamma-\log(2)\right)\epsilon \right]
 \left\{ 1 - \frac{1}{8}\left(\frac{q^2}{m\mu}\right) 
      \left(1-\frac{\epsilon}{4} \right)+ O(q^4) \right\}
    \left(\frac{m\phi_0}{2\pi}\right)^{d/2}.
\ee
We observe that $\chi(0)$ matches the NLO prediction from the
relation $\chi(0)=-(\partial n)/(\partial \mu)$, see equ.~(\ref{chi_sr}).

\section{Matching}
\label{sec_mat}

 We found that the leading terms in a small momentum expansion of the 
phonon dispersion relation and the static susceptibility satisfy the 
low energy predictions $v_s^2=(\partial P)/(\partial\rho)$ and $\chi(0)
=-(\partial n)/(\partial \mu)$ at NLO in the epsilon expansion. This 
means that we can use the curvature terms to fix the low energy 
constants $c_1$ and $c_2$. In order to be consistent with the 
low energy theorems we have to perform the matching in $d=4-\epsilon$ 
dimensions. The low energy effective Lagrangian in $d$ dimensions is
\be
\label{l_eft_d}
  {\cal L} = c_0 m^{d/2} X^{1+d/2} 
  + c_1 m^{d/2-1} \frac{(\vec{\nabla} X)^2}{X^{2-d/2}} 
  + \frac{c_2}{m^{2-d/2}} 
     \left[ \left(\nabla^2\varphi\right)^2 
           - d^2 m \nabla^2 V\right] X^{d/2-1}\,.
\ee
The powers of $m$ and $X$ follow from the scaling dimension of the 
fields. The factor $d^2$ in the $c_2$-term is a non-trivial 
consequence of conformal invariance in $d$ dimensions \cite{Son:2005rv}.
In $d$ dimensions the relation between $c_0$ and $\xi=\mu/\epsilon_F$ is 
\be
  c_0 = \frac{2}{(2\pi)^{d/2}\Gamma(2+d/2)\xi^{d/2}}\,.
\ee
The two NLO parameters $c_1,c_2$ can be related to the momentum 
dependence of the phonon dispersion relation and the static
susceptibility. In $d$ dimensions we find 
\be
\label{ph_disp_d}
 q_0 =  v_s q\left[ 1
   - \frac{4}{d(d+2)c_0}\left(c_1+\frac{d}{2} c_2\right)
                \frac{q^2}{m\mu} \right]
\ee
and
\bea
\label{chi_d}
 \chi(q) &=& - \frac{d(d+2)c_0}{4} \, m^{d/2}\mu^{d/2-1}\left[
    1 + \frac{8}{d(d+2)c_0}
    \left(c_1 - d^2\left(\frac{d}{2}-1\right) c_2\right) 
    \frac{q^2}{m\mu}  \right] \, .
\eea
We can now match the curvature terms in equ.~(\ref{ph_disp_nlo})
and (\ref{chi_nlo}) to equ.~(\ref{ph_disp_d}) and (\ref{chi_d}).
From the phonon dispersion relation we get
\be 
 c_1+\frac{d}{2}c_2=-\frac{d(d+2)c_0}{64}
  \left( 1-\frac{\epsilon}{4} \right) . 
\ee
Matching the static susceptibility gives
\be
  c_1+\frac{d}{2}c_2= c_1-d^2\left(\frac{d}{2}-1\right) c_2  \, .
\ee
This implies that $c_2$ vanishes to NLO in the epsilon expansion
$c_2/c_1=O(\epsilon^2)$. This is (barely) consistent with the 
constraint $c_2>0$ \cite{Son:2005rv}. 
The ratio $c_1/c_0$ is given by
\be 
\frac{c_1}{c_0}= -\frac{3}{8}
 \left( 1-\frac{2\epsilon}{3}+\ldots\right) . 
\ee
At NLO we obtain the following density functional for non-relativistic 
fermions at infinite scattering length 
\be
\label{final}
{\cal E}(x) = n(x)V(x) 
  + 1.364\,\frac{n(x)^{5/3}}{m}
  + 0.022\,\frac{\left(\nabla n(x)\right)^2}{mn(x)}
  + O(\nabla^4 n) \, .
\ee
It is interesting to compare this result to the density functional for 
non-interacting fermions \cite{Brack:1997}
\be
\label{ETFT}
{\cal E}_{ETF}(x) = n(x)V(x) 
  + 2.871\,\frac{n(x)^{5/3}}{m}
  + 0.014\,\frac{\left(\nabla n(x)\right)^2}{mn(x)}
  + 0.167\,\frac{\nabla^2 n(x)}{m}
  + O(\nabla^4 n) \, ,
\ee
which is known at the ``extended Thomas-Fermi model'' (ETF). The energy of 
$N$ fermions in a spherically symmetric harmonic trap is 
\be 
E=\frac{\sqrt{\xi}}{4} \omega (3N)^{4/3}
\left( 1 + \frac{c_s}{(3N)^{2/3}} + \ldots \right) \, , 
\ee
where $\xi\simeq 0.475$ (see equ.~(\ref{xi_eps})) and $c_s=-(32c_1)/(5c_0\xi)
\simeq 1.68$ at NLO in the epsilon expansion. The result for free fermions is 
\be 
E_{ETF}=\frac{1}{4} \omega (3N)^{4/3}
\left( 1 + \frac{1}{2(3N)^{2/3}} + \ldots \right) \, , 
\ee
We observe that the coefficient of the $N^{4/3}$ term in the ETF functional 
is larger than the corresponding coefficient in the unitarity limit. This 
simply reflects the fact that the interaction between the fermions is 
attractive and $\xi<1$. What is more surprising is the fact that the ETF 
functional corresponds to a significantly smaller value of $c_s$. Numerical 
results for up to $N=30$ harmonically trapped fermions can be found in 
\cite{Chang:2007,Stecher:2007,Stecher:2007b}. For small $N$ the corrections 
to the local density approximation are not very well fit by a $N^{-2/3}$ 
contribution, and the authors of \cite{Chang:2007,Stecher:2007} did not 
attempt to extract $\xi$ and $c_s$ independently. Under the assumption 
that the data can be described by $E=\xi E_{ETF}$ they find values 
$\xi\simeq (0.47-0.50)$ which are larger than the commonly accepted 
bulk value $\xi \simeq (0.40-0.44)$
\cite{Astrakharchik:2004,Carlson:2005kg,Bulgac:2006,Gezerlis:2007fs}. On 
the other hand, accepting the bulk value $\xi=(0.40-0.44)$ implies larger 
values of $c_s$ than the one predicted by the extended Thomas-Fermi model. 
Taking $E(N=20)=(41.3-43.2)\omega$ from \cite{Chang:2007,Stecher:2007} 
and $\xi=(0.40-0.44)$ gives $c_s=(0.9-2.5)$, consistent with our result 
$c_s=1.68$. We note that the data of 
\cite{Chang:2007,Stecher:2007,Stecher:2007b} are even better fit by a 
functional of the form $E(N)= \omega\sqrt{\xi}/4\cdot (3N)^{4/3}(1+
c/(3N)^{1/3})$. A correction of the form $1/N^{1/3}$ cannot be obtained 
from a local energy density functional of the form given in equ.~(\ref{edf}), 
nor is it compatible with the structure of non-leading terms in $N$ 
generally assumed in the literature \cite{Tan:2004,Bulgac:2007}. It 
is an interesting challenge to determine whether more complicated 
functionals (see Sec.~\ref{sec_out}) can yield corrections to the 
energy of $N$ harmonically trapped fermions that scale as $1/N^{1/3}$.

\section{Summary and Outlook}
\label{sec_out}

 We have computed an energy density functional for dilute non-relativistic 
fermions at unitarity. Our approach is based on an effective field theory 
of the unitary gas which takes into account the effects of spontaneous 
symmetry breaking, the existence of Goldstone modes, and the constraints 
from Galilean and conformal symmetry. We have used an epsilon expansion 
to compute the coefficients in the effective lagrangian at NLO in the 
derivative expansion. Our main result is given in equ.~(\ref{final}). 
It is interesting that at NLO in the epsilon expansion only one of the 
two possible two-derivative terms appears. 

 There are several interesting lines of investigation that we wish to 
pursue in the future. The first is the problem of constructing energy 
density functionals that depend on more than one type of density. One 
may consider, for example, the spin density (mostly of interest for 
applications in atomic physics), or the superfluid density 
\cite{Yu:2002kc,Furnstahl:2006pa,Bulgac:2007wm,Kryjevski:2007au}. 
It is also important to find a systematic way of constructing functionals 
of the Kohn-Sham type \cite{Kohn:1965,Papenbrock:2005bd}, or extensions 
of Kohn-Sham theory that contain anomalous densities (as in the 
Hartree-Fock-Bogoliubov approximation), see \cite{Yu:2002kc}. Finally, 
it is important to study more realistic interactions for neutron matter, 
in particular the effects of a finite scattering length or non-zero 
effective range \cite{Bhattacharyya:2006fg}, and the effects of 
explicit pion degrees of freedom

Acknowledgments: This work was supported by US Department of Energy 
grants DE-FG02-03ER41260. We thank A.~Kryjevski for communications
regarding his work on the Landau-Ginzburg functional \cite{Kryjevski:2007au},
as well as unpublished work on the particle number susceptibility. We 
also thank D.~Son for discussions regarding the matching procedure, 
and L.~Salasnich for comments on the manuscript.

\appendix
\section{Two-loop self energy diagrams}
\label{sec_app}

 In the appendix we compute the two-loop self energy diagrams 
that contribute to the phonon propagator at NLO. The two-loop 
``vertex-type''  diagram shown in Fig.~\ref{fig_self_nlo}b) 
is given by
\begin{align}
-i\Pi_{11}^{(b)}(p)=& -g^4 \int
 \frac{dq^d}{(2\pi)^d}\frac{dk^d}{(2\pi)^d}\frac{dq_0}{2\pi}\frac{dk_0}{2\pi}
 \[
  G_{11}(k)G_{21}(q)G_{22}(q-p)G_{12}(k-p)D(k-q)\right.\\
&\left.+G_{12}(k)G_{11}(q)G_{21}(q-p)G_{22}(k-p)D(q-k)
\], \nonumber\\
-i\Pi_{21}^{(b)}(p)=& -g^4 \int
 \frac{dq^d}{(2\pi)^d}\frac{dk^d}{(2\pi)^d}\frac{dq_0}{2\pi}\frac{dk_0}{2\pi}
\[
G_{12}(k)G_{12}(q)G_{22}(q-p)G_{11}(k-p) D(k-q)\right.\nonumber\\
&\left.+G_{22}(k)G_{11}(q) G_{12}(q-p) G_{12}(k-p) D(q-k)\]. \nonumber
\end{align}
In the $\epsilon$-expansion, counting $p_0\sim p^2\sim\epsilon$ we
only need the self-energy correction at zero energy and momentum. 
We get
\begin{align}
-i\Pi_{11}^{(b)}(0)=&-i\Pi_{21}^{(b)}(0)\\
=&-ig^4\int \frac{dq^d}{(2\pi)^d}\frac{dk^d}{(2\pi)^d}
\frac{\partial}{\partial q_0} \frac{\partial}{\partial k_0}
\[\frac{k_0+\epsilon_k}{(k_0-E_k)^2}\frac{q_0-\epsilon_k}{(q_0-E_q)^2}
\frac{1}{k_0-q_0-\epsilon_{q-k}/2}
\]\Big|_{\begin{array}{c}{q_0\rightarrow E_q}\\[-0.15cm]
  { k_0\rightarrow - E_k}\end{array}}\nonumber\\[-0.2cm]
=&-i\frac{4\phi_0\epsilon^2}{\pi} a + O(\epsilon^3)\, . \nonumber
\end{align} 
The constant $a$ can be determined by performing the integrals 
in $d=4$ spatial dimensions. A numerical calculation gives 
$a\simeq -0.267359$.

The ``self energy-type'' two-loop Feynman diagram Fig.~\ref{fig_self_nlo}c) 
is given by
\begin{align}
-i\Pi_{11}^{(c)}=&-2 g^4\int
 \frac{dq^d}{(2\pi)^d}\frac{dk^d}{(2\pi)^d}\frac{dq_0}{2\pi}\frac{dk_0}{2\pi}
\[G_{11}(k)
G_{11}(k)G_{22}(q)G_{22}(k-p)D(k-q)\right.\\
&\left.+G_{11}(q)G_{12}(k)G_{21}(k)G_{22}(k-p)D(q-k)\]\equiv
B+F\, ,\nonumber\\
-i\Pi_{21}^{(c)}=&-2g^4
\int
 \frac{dq^d}{(2\pi)^d}\frac{dk^d}{(2\pi)^d}\frac{dq_0}{2\pi}\frac{dk_0}{2\pi}
\[G_{11}(k)G_{21}(k)G_{21}(k-p)G_{22}(q)D(k-q)\right.\nonumber\\
&\left.+G_{22}(k)G_{21}(k)G_{21}(k-p)G_{11}(q) D(q-k)\]\equiv H+K
 \, .\nonumber
\end{align} 
In the limit $p=0$, changing variables $k_0\rightarrow -k_0$ shows that
$F=K$ and $F=H$. We get
\begin{align}
B=&-ig^4\int \frac{dq^d}{(2\pi)^d}\frac{dk^d}{(2\pi)^d}
\frac{E_q-\epsilon_q}{2E_q}\frac{\partial^2}{\partial k_0^2}
\[\frac{(k_0+\epsilon_k)^2(k_0-\epsilon_k)}
  {(k_0-E_k)^3(k_0-E_q-\epsilon_{k-q}/2)}
\]\Big|_{k_0=-E_k}\\
\equiv & -i\frac{2\phi_0\epsilon^2}{\pi}b\, , \nonumber\\
F= &-ig^4\int \frac{dq^d}{(2\pi)^d}\frac{dk^d}{(2\pi)^d}
\frac{E_q-\epsilon_q}{2E_q}\frac{\partial^2}{\partial k_0^2}
\[\frac{(k_0-\epsilon_k)\phi_0^2}{(k_0+E_k)^3(-k_0-E_q-\epsilon_{k-q}/2)}
\]\Big|_{k_0=E_k}\nonumber\\
\equiv& -i\frac{2\phi_0\epsilon^2}{\pi} f \, .\nonumber
\end{align}
Numerical evaluation gives $b\simeq -0.5822930$ and $f=h=k\simeq
0.09742858$. Collecting all the terms
\begin{align}\label{numC12}
 \Pi_{11}^{(b)}(0) + \Pi_{11}^{(c)}(0)
 &=  \frac{2 \phi_0\epsilon^2}{\pi}(2a+b+f)
  =  C_1\phi_0\epsilon^2
  \simeq -0.64908 \phi_0\epsilon^2 \, , \\
 \Pi_{21}^{(b)}(0)+\Pi_{21}^{(c)}(0)
 &=  \frac{4\phi_0\epsilon^2}{\pi}(a+f) 
  =  C_2\phi_0\epsilon^2
  \simeq -0.21636 \phi_0\epsilon^2 \, ,\nonumber
\end{align}
which agree very well with the determination in equ.~(\ref{C_12}).


\end{document}